\documentclass[showpacs,showkeys,onecolumn,pr,preprintnumbers,amsmath,amssymb]{revtex4}
\newcommand{\beq}{\begin{equation}}
\newcommand{\eeq}{\end{equation}}

\newcommand{\id}{i\kern.06em\hbox{\raise.25ex\hbox{$/$}\kern-.60em$\partial$}}

\newcommand{\rD}{\!\not\!\! D}
\newcommand{\bs}{/\kern-.52em b}
\newcommand{\qs}{/\kern-.52em s}

\newcommand{\p}{\partial}
\newcommand{\yp}{^{\prime}}

\newcommand{\dd}{\kern.06em\hbox{\raise.25ex\hbox{$/$}\kern-.60em$\partial$}}
\newcommand{\gl}{\kern.06em\hbox{\hbox{$\langle\,$}\kern-.60em$\,\langle$}} 
\newcommand{\gr}{\kern.06em\hbox{\hbox{$\rangle\,$}\kern-.60em$\,\rangle$}}

\newcommand{\vep}{\varepsilon}
\newcommand{\Tr}{\mathop{\rm Tr}\nolimits}

\newcommand{\ds}{/\kern-.52em d}

\newcommand{\bj}{{\boldsymbol j}}

\newcommand{\bk}{{\boldsymbol k}}
\newcommand{\bp}{{\boldsymbol p}}

\newcommand{\br}{{\boldsymbol r}}

\newcommand{\ms}{\boldsymbol{s}}
\newcommand{\ma}{\boldsymbol{a}}

\newcommand{\mr}{\boldsymbol{r}}

\newcommand{\bsigma}{\boldsymbol{\sigma}}
\newcommand{\bgamma}{\boldsymbol{\gamma}}

\newcommand{\sumn}{\sum\limits}
\newcommand{\balpha}{\boldsymbol{\alpha}}

\newcommand{\bfKpm}{{\boldsymbol{K}_{\pm}}}
\newcommand{\bKp}{{\boldsymbol{K}_+}}
\newcommand{\bKm}{{\boldsymbol{K}_-}}

\newcommand{\bA}{{\boldsymbol A}}
\newcommand{\bB}{{\boldsymbol B}}

\newcommand{\bx}{\boldsymbol{x}}
\newcommand{\bv}{{\boldsymbol v}}
\newcommand{\bE}{{\boldsymbol E}}
\newcommand{\Bohrmu}{\mu_{_{\rm B}}}
\newcommand{\Fermiv}{v_{_{\rm F}}}

\newcommand{\lj}{\langle}
\newcommand{\rj}{\rangle}

\DeclareMathAlphabet{\mathpzc}{OT1}{pzc}{m}{it}
\usepackage{mathrsfs}
\input amssym.def
\input amssym
\input epsf
\usepackage{graphicx}
\usepackage{dcolumn}
\usepackage{bm}
\usepackage{epsfig}

\begin{document}
\title{Graphene in external fields}
\author{Sze-Shiang Feng }
\affiliation
{ Physics Department, Wright State  University, Dayton, OH 45435}
\email{ sze-shiang.feng@wright.edu}
\author{Mogus Mochena}
\affiliation{ Physics Department, Florida A \& M  University, Tallahassee, FL 32307}
\date{\today}
\begin{abstract}
A general discussion of graphene in external electromagnetic field is provided. In general, the formulation is not Lorentz
invariant because of Zeeman energy. But it can be restored approxiamtely in the case of strong magnetic field, the condition when quantum Hall effect is observed.  Besides obtaining the well-known Hall conductance $\frac{4q^2}{h}(L+1/2)$, we also provide an explanation of
the newly observed Hall conductance $\frac{4q^2}{h}L$ for $L=0,1$. These are part of the
sequence of Hall conductance $\frac{4q^2}{h}L$ which depends on the filling of Zeeman levels. The energy levels are obtained for general orthogonal constant electric and magnetic field. The second order Dirac equation is derived in applied monochromatic electromagnetic wave, and the major difference between graphene system and convetional Dirac electrons is pointed out.
\end{abstract}
\maketitle
\vskip 0.3in
\section{Introduction}
Graphene has been one of the major foci in physics because of its simple lattice structure and linear dispersion relation near the Fermi level \cite{Neto:2009}-\cite{Gusynin:2008} when only nearest-neighbor hopping is taken into account. It has become a new testbed not only for condensed matter physics, but also for quantum field theory and mathematical physics \cite{Katsnelson:2007}-\cite{Mikitik:2008}. The physical properties of graphene in external field such as quantum Hall effect (both integer and fractional \cite{Toke:2006}-
\cite{Gorbar:2008}),
 spin quantum Hall effect \cite{Kane:2005}, transport theory \cite{Hwang:2007}\cite{Yan:2008}, superconducting\cite{Jiang:2008}, and magnetic confinement\cite{Martino:2007}
 are under intensive study. Effects of next-nearest-neighbor hopping have also been studied \cite{Suprunenko:2008}.
  It is widely recognized that the integer quantum Hall conductance is $\sigma_{xy}=\frac{4e^2}{h}(n+\frac{1}{2})$ where $n=0,1,\cdots$\cite{Neto:2009}.
  Although disorder and 4-fold symmetry breaking may be utilized to explain the newly found quantum Hall structures $\nu=0,\pm 1, \pm 4$\cite{Fuchs:2007} 
-\cite{Jiang:2007}, a unified explanation is still called for and the simple structure of $\sigma_{xy}=\frac{4e^2}{h}(n+\frac{1}{2})$ still deserves a simple and fundamental explanation.  Not surprisingly, quantum Hall plateaus $\nu=\pm 2,\pm 6,\pm 10$ in graphenes can be explained by using the Landau levels of spinless particle in external magnetic field and the 1+2 Lorentz invariance of the massless Dirac Hamiltonian \cite{Beneventano:2007a}. Because simple utilisation of Landau levels does not consider the Zeeman energy, which is not negligible compared with low-lying Landau levels, Zeeman energy might explain some of the newly found plateaus. Yet, Zeeman energy is not Lorentz invariant. Therefore, how to use the well-known Landau levels and the Lorentz transformation property of Dirac Hamiltonian remains an issue.  In this paper, we discuss graphene in general external fields. Because of the Zeeman energy, the system is not SO(1,2) invariant in general. But when the magnetic field is strong enough, the system does enjoy the 
invariance approximately, which can be utilized to relate the physics of one experimental configuration to that of another. We obtain the energy levels of the system for the case of  constant magnetic fields, and for constant electric and magnetic field.  Finally, we derive the second-order Dirac equation in monochromatic electromagnetic waves and point out the difference bewteen graphene system and conventional relativistic electrons. The rest of the paper is organized as follows: in Section II, we present the Hamiltonian in general applied field in the nearest-neighbor hopping approximation and discuss condition of SO(1,2) invariance of the corresponding Lagrangian. In Section III, we discuss the energy levels of the Hamiltonian in applied perpendicular magnetic field. We dicsuss in Section IV the energy levels in applied orthogonal magnetic and electric fields and derive the quantum Hall conductance in the case of strong magnetic field using the approximate SO(1,2) invariance. A discussion of Dirac equation in applied monochromatic electromagnetic wave is also provided using the method of Volkow in this section. The last section is a brief summary.
\section{Hamiltonian in applied field}
\indent The direct lattice of graphene is a superposition of two interpenetrated triangular lattices $\Lambda_A,  \Lambda_B$. The generators of lattice $\Lambda_A$ are \cite{Beneventano:2007} $\ma_1=\sqrt{3}a(\frac12,\frac{-\sqrt{3}}{2})$, and $\ma_2=\sqrt{3}a(\frac12,\frac{\sqrt{3}}{2})$, where $a\approx 1.42$ \AA\, is the carbon-carbon distance. The vectors $\ms_1=a(0,-1)$, $\ms_2=a(\frac{\sqrt{3}}{2},\frac12)$ and $\ms_3=a(\frac{-\sqrt{3}}{2},\frac12)$ connect each site in the lattice $\Lambda_A$ to its nearest neighbor sites in the lattice $\Lambda_B$.
Unlike regular electron spin, the pseudospin in graphene represents the two sublattices and there is no magnetic moment associated hence does not couple directly
to magnetic field\cite{Toke:2006}.
 The tight binding Hamiltonian can then be written 
 \beq
 H_0=-t\sum_{\sigma}\sum_{\mr\in \Lambda_A}\sum^{3}_{i=1}\left[ A_\sigma^\dagger(\mr)B_\sigma(\mr+\ms_i)+B_\sigma^\dagger(\mr+\ms_i)
    A_\sigma(\mr)\right]
 \eeq
where $\sigma$ is pseudo-spin index and $t$ is the uniform hopping constant. In the presence of applied magnetic field $\bB$ and electric fields $\bE=-\nabla\varphi$, Zeeman energy and Coulomb energy should be included.
\begin{eqnarray}
H_Z&=&\mu_{_{\rm B}}[\sum_{\boldsymbol{r}\in\Lambda_A}\boldsymbol{B}\cdot A^\dag(\boldsymbol{r})\boldsymbol{\tau}A(\boldsymbol{r})+
\sum_{\boldsymbol{r}\in\Lambda_B}\boldsymbol{B}\cdot B^\dag(\boldsymbol{r})\boldsymbol{\tau}B(\boldsymbol{r})]
\end{eqnarray}
\beq
H_C=q[\sum_{\boldsymbol{r}\in\Lambda_A}\varphi(\br)A^\dag(\boldsymbol{r})A(\boldsymbol{r})+
\sum_{\boldsymbol{r}\in\Lambda_B}\varphi(\br) B^\dag(\boldsymbol{r})B(\boldsymbol{r})]
\eeq
, $\mu_{_{\rm B}}=\frac{|e|\hbar}{2m_e}$ is the Bohr magneton. 
In momentum space, with
$a_\bk=\frac{1}{\sqrt{N_\Lambda}}\sumn_{\mr \in\Lambda_A}e^{-i \bk \cdot \br}A(\br),
b_\bk=\frac{1}{\sqrt{N}_\Lambda}\sumn_{\mr \in\Lambda_B}e^{-i \bk \cdot \br}B(\br),$
where $N_\Lambda$ is the number of lattice points of sublattice $\Lambda_A$ (or $\Lambda_B$), $H_0+H_Z$ reads
in $\bk$-space
\beq
 H=-t\sum_\bk (f_\bk a_{\sigma\bk}^\dag b_{\sigma\bk}+f^\ast_{\bk} b^\dag_{\sigma\bk} a_{\sigma\bk})
 +\mu_{\rm B}\boldsymbol{B}\cdot\sum_\bk (a_\bk^\dag\boldsymbol{\tau}a_\bk+b_\bk^\dag\boldsymbol{\tau}b_\bk)
  \eeq
with $f_\bk = \sum_{i=1}^{3} e^{i \bk \cdot \ms_i }$. $t\approx 3.033$ eV.
\beq
f_\bk=e^{-ik_ya}+2e^{\frac{i}{2}k_ya}\cos(\frac{\sqrt{3}}{2}k_xa)
\eeq
The lattice Hamitonian $H_0$ vanishes at the six corners of the first Brillouin zone. Among these, only two are inequivalent, and can be chosen as
\beq
\bfKpm = \pm \left(\frac{4\pi}{3\sqrt{3}\,a}, 0\right) ;f_{\bfKpm} = 0.
\eeq
Let $\bk = \bfKpm +\bp/\hbar$.
In the vicinity of  ${\bf K}_+$, we have
\beq
f_\bk=-\frac{3}{2}p_xa/\hbar-\frac{3}{2}ip_ya/\hbar+\frac{3}{8}a^2(p_x-ip_y)^2/\hbar^2+\cdots
\eeq
In the vicinity of  ${\bf K}_-$, we have
\beq
f_\bk=\frac{3}{2}p_xa/\hbar-\frac{3}{2}ip_ya/\hbar+\frac{3}{8}a^2(p_x+ip_y)^2/\hbar^2+\cdots
\eeq
After defining two-component spinors as
$\psi_\bk=(a_\bk, b_\bk)^T, \psi^\dag_\bk=(a^\dag _\bk, b^\dag_\bk),\psi_{\bk\sigma 1}=a_{\bk\sigma},
\psi_{\bk\sigma 2}=b_{\bk\sigma},\sigma=\pm 1$.
we have
\beq
H=-t\sum_{\bk\sigma}\psi^\dag_{\bk\sigma} \left(
      \begin{array}{cc}
        0 & f_\bk \\
       f^\ast_\bk & 0 \\
      \end{array}
    \right)\psi_{\bk\sigma}+\mu_{\rm B}\boldsymbol{B}\cdot\sum_\bk (\psi_{\bk\alpha 1}^\dag\bsigma_{\alpha\beta}\psi_{\bk\beta 1}+\psi_{\bk\alpha 2}^\dag\bsigma_{\alpha\beta}\psi_{\bk\beta 2})
    +H_C
    \eeq
When $H_0$ is linearized around these two points , one obtains
\beq
H_{0|\bk=\bKp+\bp}=
v_{_{\rm F}}\left(\begin{array}{cc}
        0 & p_x+ i p_y \\
        p_x-i p_y & 0 \\
      \end{array}\right)=v_{_{\rm F}}\balpha\cdot\bp
      \eeq
where $v_{_{\rm F}} =\frac{3ta}{2\hbar},\balpha=(\sigma_x,-\sigma_y)$. 
and
\beq
H_{0|\bk=\bKm+\bp}=
-v_{_{\rm F}}\left(\begin{array}{cc}
        0 & p_x- i p_y \\
        p_x+i p_y & 0 \\
      \end{array}\right)=-v_{_{\rm F}}\bsigma\cdot\bp
      \eeq
$   \bsigma=(\sigma_x,\sigma_y)$ .
In the case $\boldsymbol{B}=(0,0,B)$, we have
\beq
H=-t\sum_{\bk\sigma}\psi^\dag_{\bk\sigma} \left(
      \begin{array}{cc}
        0 & f_\bk \\
       f^\ast_\bk & 0 \\
      \end{array}
    \right)\psi_{\bk\sigma}+\mu_{_{\rm B}}B\sum_{\bk}(\psi^\dag_{\bk +}\psi_{\bk +}-\psi^\dag_{\bk -}\psi_{\bk -})
    +H_C
    \eeq
Or, let $\psi_{\bk+}=\xi_\bk=\left(\begin{array}{c}a_{\bk+}\\b_{\bk+}\end{array}\right),\psi_{\bk-}=\eta_\bk
=\left(\begin{array}{c}a_{\bk-}\\b_{\bk-}\end{array}\right)
$, we have
\beq
H=-t\sum_{\bk}\xi^\dag_{\bk} \left(
      \begin{array}{cc}
        0 & f_\bk \\
       f^\ast_\bk & 0 \\
      \end{array}
    \right)\xi_{\bk}
    -t\sum_{\bk}\eta^\dag_{\bk} \left(
      \begin{array}{cc}
        0 & f_\bk \\
       f^\ast_\bk & 0 \\
      \end{array}
    \right)      \eta_{\bk}
        +\mu_{_{\rm B}}B\sum_{\bk}(\xi^\dag_{\bk}\xi_{\bk}-\eta^\dag_{\bk}\eta_{\bk})+H_C
    \eeq
Close to  ${\bf K_+}$, we have
\beq
H=\sum_{\bp}\xi^\dag_{\bp} (v_{_{\rm F}}\balpha\cdot\bp+\mu_{_{\rm B}}B)\xi_{\bp}+
    \sum_{\bp}\eta^\dag_{\bp}(v_{_{\rm F}} \balpha\cdot\bp-\mu_{_{\rm B}}B)\eta_{\bp}+H_C
\eeq
There are two inequivalent representations of the $\gamma$-matrices
in three dimensions:$\gamma^{\mu}=(\sigma_3, i\sigma_2, i\sigma_1)$
and $\gamma^{\mu}=(-\sigma_3, -i\sigma_2, -i\sigma_1)$. We choose the first. 
\beq
H=\sum_{\bp}\bar{\xi}_{\bp} ( v_{_{\rm F}}\bgamma\cdot\bp+\mu_{_{\rm B}}B\gamma^0)\xi_{\bp}+
    \sum_{\bp}\bar{\eta}_{\bp}( v_{_{\rm F}} \bgamma\cdot\bp-\mu_{_{\rm B}}B\gamma^0)\eta_{\bp}+H_C
\eeq
where $\bar{\xi}=\xi^\dag\gamma^0.$ 
Incorporated $U(1)$ gauge invariance, the Hamiltonian reads
\beq
H=\int d^2\bx\bar{\xi}(x) [\hbar  v_{_{\rm F}}\bgamma\cdot(-i\nabla-\frac{q}{\hbar}\bA)+\mu_{_{\rm B}}B\gamma^0+q\gamma^0\varphi]\xi(x)+
    \int d^2\bx\bar{\eta}(x)[\hbar   v_{_{\rm F}} \bgamma\cdot(-i\nabla-\frac{q}{\hbar}\bA)-\mu_{_{\rm B}}B\gamma^0+q\gamma^0\varphi]\eta(x)
\eeq
Let $x^\mu=(x^0,\boldsymbol{x})=(v_{_{\rm F}}t,x,y), A^\mu=(A^0, \bA)$.
Denoting $D_\mu=\p_\mu+i\frac{q}{\hbar}A_\mu, A^0=\varphi/v_{_{\rm F}}$, we have
\beq
\mathcal{L}=\xi^\dag i\hbar \p_t\xi-\mathcal{H}=
\bar{\xi}(x) (i\hbar\rD+gB\gamma^0)\xi(x)+\bar{\eta}(x)(i\hbar\rD-gB\gamma^0)\eta(x)
\eeq
where $g=\mu_{_{\rm B}}/v_{_{\rm F}}$. The U(1) gauge invariance is preserved but the Lorentz SO(1,2) invariance is broken by the Zeeman term. 
Since
\beq
-\vep^{\mu\nu\tau}\p_\mu A_\nu \gamma_\tau=B\gamma^0+\frac{1}{\Fermiv}E_x\gamma^2-\frac{1}{\Fermiv}E_y\gamma^1
\eeq
For $|\bE| <<\Fermiv |\bB|$
, we can write
\beq
B\gamma^0\approx-\vep^{\mu\nu\tau}\p_\mu A_\nu \gamma_\tau
\eeq
and in this case
\beq
L=\int d^3x \Big[\bar{\xi}(x) (i\hbar\rD-g\vep^{\mu\nu\tau}\p_\mu A_\nu \gamma_\tau)\xi(x)+\bar{\eta}(x)(i\hbar\rD+g\vep^{\mu\nu\tau}\p_\mu A_\nu \gamma_\tau)\eta(x)\Big]
\eeq
which shows 1+2 Lorentz invariance. The current is
\beq
j^\mu=-\frac{\delta L}{\delta A_\mu}=q\bar{\xi}\gamma^\mu\xi+q\bar{\eta}\gamma^\mu\eta+g
\vep^{\mu\nu\tau}\p_\nu(\bar{\xi}\gamma_\tau\xi)
-g\vep^{\mu\nu\tau}\p_\nu(\bar{\eta}\gamma_\tau\eta)
\eeq
\section{Constant perpendicular magnetic field}
In this case $\varphi=0$  and we want to calculate the grand-canonical partition function.$Z=\Tr e^{-\beta K}, K=H-\mu N$.
The Dirac equation is
\beq
\Big[\hbar v_{_{\rm F}}\balpha\cdot(-i\nabla-\frac{q}{\hbar}\bA)+\mu_{_{\rm B}}B-\mu\Big]\xi=K\xi
\eeq
which turns into second order
\beq
(-D_iD_i+\frac{q}{\hbar}B\sigma_3)\xi=\frac{(K-\mu_{_{\rm B}}B+\mu)^2}{\hbar^2v_{_{\rm F}}^2}\xi
\eeq
Using standard Landau levels (assuming $qB=|qB|$), we have
\beq
(K-\mu_{_{\rm B}}B+\mu)^2=\hbar v_{_{\rm F}}^2 2qB(\ell+1/2-s_z)
\eeq
So
\beq
K_{\ell,s_z}=\pm \hbar v_{_{\rm F}}\sqrt{2\frac{q}{\hbar}B(\ell+1/2-s_z)}+\mu_{_{\rm B}}B-\mu
\eeq
The partition function is
\beq
Z=\prod_{\ell,s_z}[1+e^{-\beta K_{\ell,s_z}}]^{\Delta_{_{\rm L}}}
\eeq
where $\Delta_{_{\rm L}}=\frac{|qB|}{2\pi\hbar}$ is the Landau degeneracy per unit area. For $B=1$T, we have
$\Delta_{_{\rm L}}\approx 2.4\times 10^{14}/m^2$.
\beq
-\Gamma=\Delta_{_{\rm L}}\sum_\ell\sum_{s_z}\ln [1+e^{-\beta K_{\ell,s_z}}]\label{Z}
\eeq
For $\mu=0$, we have the energy levels
\beq
K_{\ell,s_z}=\pm \hbar v_{_{\rm F}}\sqrt{2\frac{q}{\hbar}B(\ell+1/2-s_z)}+\mu_{\rm B}B
\eeq
Corresponding to field $\eta$, we have
\beq
K_{\ell,s_z}=\pm \hbar v_{_{\rm F}}\sqrt{2\frac{q}{\hbar}B(\ell+1/2-s_z)}-\mu_{\rm B}B
\eeq
So the energy levels are symmetric under $+\leftrightarrow -$. Therefore if the graphene is undoped, the Fermi level is still at $\mu=0$. Suppose $\mu_{\rm B}B>0$. 
For $\xi$-field, $K_{\ell,1/2}$ levels are ($[A]$ represents integer part of $A$)
\beq
s_z=1/2: \left\{\begin{array}{cl}    
\mu_{\rm B}B-\hbar|v_{_{\rm F}}|\sqrt{2\frac{q}{\hbar}B\ell}, & \ell=0,1,\cdots, [\frac{\mu_{\rm B}^2B}{2q \hbar v_{_{\rm F}}^2}]
 , {\rm positive }\\
 \mu_{_{\rm B}}B-\hbar |v_{_{\rm F}}|\sqrt{2\frac{q}{\hbar}B\ell}, & \ell=[\frac{\mu_{_{\rm B}}^2B}{2q\hbar v_{_{\rm F}}^2}]+1,\cdots, \infty
 , {\rm negative }\\
\mu_{_{\rm B}}B+\hbar|v_{_{\rm F}}|\sqrt{2\frac{q}{\hbar}B\ell}, & \ell=1,2,\cdots,\infty, {\rm positive}
\end{array}\right.
\eeq
$K_{\ell,-1/2}$ levels are
\beq
s_z=-1/2: \left\{\begin{array}{cl}    
\mu_{_{\rm B}}B-\hbar|v_{_{\rm F}}|\sqrt{2\frac{q}{\hbar}B(\ell+1)}, & \ell=0, 1,\cdots, [\frac{\mu_{_{\rm B}}^2B}{2q\hbar v_{_{\rm F}}^2}]-1
 ,{\rm positive}\\
 \mu_{_{\rm B}}B-\hbar|v_{_{\rm F}}|\sqrt{2\frac{q}{\hbar}B(\ell+1)}, & \ell=[\frac{\mu_{_{\rm B}}^2B}{2q\hbar v_{_{\rm F}}^2}],\cdots,\infty  ,{\rm negative}\\
\mu_{_{\rm B}}B+\hbar|v_{_{\rm F}}|\sqrt{2\frac{q}{\hbar}B(\ell+1)}, & \ell=0,1,2,\cdots\infty, {\rm positive}
\end{array}\right.
\eeq
For $\eta$-field, $K_{\ell,1/2}$ levels are
\beq
s_z=1/2: \left\{\begin{array}{cl}    
-\mu_{_{\rm B}}B+\hbar|v_{_{\rm F}}|\sqrt{2\frac{q}{\hbar}B\ell}, & \ell=0,1,\cdots, [\frac{\mu_{_{\rm B}}^2B}{2q\hbar v_{_{\rm F}}^2}]
 , {\rm negativee }\\
 -\mu_{_{\rm B}}B+\hbar|v_{_{\rm F}}|\sqrt{2\frac{q}{\hbar}B\ell}, & \ell=[\frac{\mu_{_{\rm B}}^2B}{2q\hbar v_{_{\rm F}}^2}]+1,\cdots, \infty
 , {\rm positive }\\
-\mu_{_{\rm B}}B-\hbar|v_{_{\rm F}}|\sqrt{2\frac{q}{\hbar}B\ell}, & \ell=1,2,\cdots,\infty, {\rm negative}
\end{array}\right.
\eeq
$K_{\ell,-1/2}$ levels are
\beq
s_z=-1/2: \left\{\begin{array}{cl}    
-\mu_{_{\rm B}}B+\hbar|v_{_{\rm F}}|\sqrt{2\frac{q}{\hbar}B(\ell+1)}, & \ell=0, 1,\cdots, [\frac{\mu_{_{\rm B}}^2B}{2q\hbar v_{_{\rm F}}^2}]-1
 ,{\rm negative}\\
 -\mu_{_{\rm B}}B+\hbar|v_{_{\rm F}}|\sqrt{2\frac{q}{\hbar}B(\ell+1)}, & \ell=[\frac{\mu_{_{\rm B}}^2B}{2q\hbar v_{_{\rm F}}^2}],\cdots,\infty
 ,{\rm positive}\\
-\mu_{_{\rm B}}B-\hbar|v_{_{\rm F}}|\sqrt{2\frac{q}{\hbar}B(\ell+1)}, & \ell=0,1,2,\cdots\infty, {\rm negative}
\end{array}\right.
\eeq
So negative levels comes from
both $\xi$ and $\eta$ fields.
\beq
\xi:  K_{\ell,1/2}:\mu_{_{\rm B}}B-\hbar |v_{_{\rm F}}|\sqrt{2\frac{q}{\hbar}B\ell},  \ell=[\frac{\mu_{_{\rm B}}^2B}{2q\hbar v_{_{\rm F}}^2}]+1,\cdots, \infty
\eeq
\beq
\xi:  K_{\ell,-1/2}:\mu_{_{\rm B}}B-\hbar|v_{_{\rm F}}|\sqrt{2\frac{q}{\hbar}B(\ell+1)},  \ell=[\frac{\mu_{_{\rm B}}^2B}{2q\hbar v_{_{\rm F}}^2}],\cdots, \infty
\eeq
\beq
\eta: K_{\ell,1/2}: \left\{\begin{array}{cl}    
-\mu_{_{\rm B}}B+\hbar|v_{_{\rm F}}|\sqrt{2\frac{q}{\hbar}B\ell}, & \ell=0,1,\cdots, [\frac{\mu_{_{\rm B}}^2B}{2q\hbar v_{_{\rm F}}^2}]
 \\
 -\mu_{_{\rm B}}B-\hbar|v_{_{\rm F}}|\sqrt{2\frac{q}{\hbar}B\ell}, & \ell=1,2,\cdots,\infty
\end{array}\right.
\eeq
\beq
\eta: K_{\ell,-1/2}: \left\{\begin{array}{cl}    
-\mu_{_{\rm B}}B+\hbar|v_{_{\rm F}}|\sqrt{2\frac{q}{\hbar}B(\ell+1)}, & \ell=0, 1,\cdots, [\frac{\mu_{_{\rm B}}^2B}{2q\hbar v_{_{\rm F}}^2}]-1
 \\
 -\mu_{_{\rm B}}B-\hbar|v_{_{\rm F}}|\sqrt{2\frac{q}{\hbar}B(\ell+1)}, & \ell=0,1,2,\cdots\infty 
\end{array}\right.
\eeq
For $t=3.033$ eV, $v_{_{\rm F}}=\frac{3ta}{2\hbar}\approx  10^{6}$(m/s).  Defining
$B_0:=\frac{2qv_{_{\rm F}}^2\hbar}{\mu_{_{\rm B}}^2}$, then
$B_0\approx 1.1\times 10^6$T. For lab field $B\sim 10$T, hence, $[\frac{B}{B_0}]=0$. Thus for $\xi$-field, $K_{\ell,1/2}$ levels are
\beq
s_z=1/2: \left\{\begin{array}{cl}    
\mu_{_{\rm B}}B & \ell=0,
  {\rm positive }\\
 \mu_{_{\rm B}}B-\hbar |v_{_{\rm F}}|\sqrt{2\frac{q}{\hbar}B\ell}, & \ell=1,\cdots, \infty
 , {\rm negative }\\
\mu_{_{\rm B}}B+\hbar|v_{_{\rm F}}|\sqrt{2\frac{q}{\hbar}B\ell}, & \ell=1,2,\cdots,\infty, {\rm positive}
\end{array}\right.
\eeq
$K_{\ell,-1/2}$ levels are
\beq
s_z=-1/2: \left\{\begin{array}{cl}    
 \mu_{_{\rm B}}B-\hbar|v_{_{\rm F}}|\sqrt{2\frac{q}{\hbar}B(\ell+1)}, & \ell=0,\cdots,\infty  ,{\rm negative}\\
\mu_{_{\rm B}}B+\hbar|v_{_{\rm F}}|\sqrt{2\frac{q}{\hbar}B(\ell+1)}, & \ell=0,1,2,\cdots\infty, {\rm positive}
\end{array}\right.
\eeq
For $\eta$-field, $K_{\ell,1/2}$ levels are
\beq
s_z=1/2: \left\{\begin{array}{cl}    
-\mu_{_{\rm B}}B, & \ell=0
 , {\rm negativee }\\
 -\mu_{_{\rm B}}B+\hbar|v_{_{\rm F}}|\sqrt{2\frac{q}{\hbar}B\ell}, & \ell=1,\cdots, \infty
 , {\rm positive }\\
-\mu_{_{\rm B}}B-\hbar|v_{_{\rm F}}|\sqrt{2\frac{q}{\hbar}B\ell}, & \ell=1,2,\cdots,\infty, {\rm negative}
\end{array}\right.
\eeq
$K_{\ell,-1/2}$ levels are
\beq
s_z=-1/2: \left\{\begin{array}{cl}    
 -\mu_{_{\rm B}}B+\hbar|v_{_{\rm F}}|\sqrt{2\frac{q}{\hbar}B(\ell+1)}, & \ell=0,\cdots,\infty
 ,{\rm positive}\\
-\mu_{_{\rm B}}B-\hbar|v_{_{\rm F}}|\sqrt{2\frac{q}{\hbar}B(\ell+1)}, & \ell=0,1,2,\cdots\infty, {\rm negative}
\end{array}\right.
\eeq
Note $s_z=1/2$ belongs to sublattice $\Lambda_A$ and $s_z=-1/2$ belongs to sublattice $\Lambda_B$. Consider negative levels\\
For $\xi$-field, $K_{\ell,1/2}$ levels are
\beq
s_z=1/2: \left\{\begin{array}{cl}    
 \mu_{_{\rm B}}B-\hbar |v_{_{\rm F}}|\sqrt{2\frac{q}{\hbar}B\ell}, & \ell=1,\cdots, \infty
 , {\rm negative }\\
\end{array}\right.
\eeq
$K_{\ell,-1/2}$ levels are
\beq
s_z=-1/2: \left\{\begin{array}{cl}    
 \mu_{_{\rm B}}B-\hbar|v_{_{\rm F}}|\sqrt{2\frac{q}{\hbar}B(\ell+1)}, & \ell=0,\cdots,\infty  ,{\rm negative}\\
\end{array}\right.
\eeq
For $\eta$-field, $K_{\ell,1/2}$ levels are
\beq
s_z=1/2: \left\{\begin{array}{cl}    
-\mu_{_{\rm B}}B, & \ell=0
 , {\rm negativee }\\
 -\mu_{_{\rm B}}B-\hbar|v_{_{\rm F}}|\sqrt{2\frac{q}{\hbar}B\ell}, & \ell=1,2,\cdots,\infty, {\rm negative}
\end{array}\right.
\eeq
$K_{\ell,-1/2}$ levels are
\beq
s_z=-1/2: \left\{\begin{array}{cl}    
-\mu_{_{\rm B}}B-\hbar|v_{_{\rm F}}|\sqrt{2\frac{q}{\hbar}B(\ell+1)}, & \ell=0,1,2,\cdots\infty, {\rm negative}
\end{array}\right.
\eeq
So levels
$$ \mu_{_{\rm B}}B-\hbar |v_{_{\rm F}}|\sqrt{2\frac{q}{\hbar}B\ell},  \ell=1,\cdots, \infty$$
 and levels
$$-\mu_{_{\rm B}}B-\hbar|v_{_{\rm F}}|\sqrt{2\frac{q}{\hbar}B\ell},  \ell=1,2,\cdots,\infty$$
are doubly degenerate(apart from the Landau degeneracy). We can not have an infinite number of negative levels filled-up since a cut-off is necessary. Suppose among the levels  
$\mu_{_{\rm B}}B-\hbar|v_{_{\rm F}}|\sqrt{2\frac{q}{\hbar}B\ell},  \ell=L_1^u\cdots, L_1^l$ are filled, and among the levels $-\mu_{_{\rm B}}B-\hbar|v_{_{\rm F}}|\sqrt{2\frac{q}{\hbar}B\ell}, \ell=L_2^u,\cdots L_2^l$ are filled. If $-\mu_{_{\rm B}}B$ is also filled, then at $T=0K$
\beq
N=\Delta_{_{\rm L}}(2(L_1^l-L_1^u+1)+2(L_2^l-L_2^u+1)+1)=2\Delta_{_{\rm L}}(L_1+L_2+\frac{1}{2})
\eeq
where $L_1=L_1^l-L_1^u+1, L_2=L_2^l-L_2^u+1$. If $-\mu_{_{\rm B}}B$ is not filled, i.e., $\mu<-\mu_{_{\rm B}}B$,
then
\beq
N=\Delta_{_{\rm L}}(2(L_1^l-L_1^u+1)+2(L_2^l-L_2^u+1))=2\Delta_{_{\rm L}}(L_1+L_2)
\eeq
The physics near $\bKm$ makes equal contribution. 
The magnetization is 
\beq
M=\mu_{_{\rm B}}(N_\xi-N_\eta)
\eeq
where
\beq
N_\xi=\sum_\bp\lj \xi^\dag_\bp\xi_\bp\rj,\,\,\,\,\,\,\,\,\,
N_\eta=\sum_\bp\lj \eta^\dag_\bp\eta_\bp\rj,\,\,\,\,\,\,\,\,\,
\eeq
The effective potential for $\xi$-fields
 should be regularized. The sum in the second term in the fowllowing formula actually extends to only $L_1$ instead of to infinity. Hence
\begin{eqnarray}
-\frac{\Gamma_\xi}{\Delta_{_{\rm L}}}
&=&
\ln\Big[1+e^{\beta(\mu-\mu_{_{\rm B}}B)}\Big]
+2\sum^\infty_{\ell=1}\ln\Big[
1+e^{\beta(\mu-\hbar v_{_{\rm F}}\sqrt{2\frac{q}{\hbar}B\ell}-\mu_{_{\rm B}}B)}
\Big]
+2\sum_{\ell=1}^{L_1}\ln
\Big[
1+e^{\beta(\mu+\hbar v_{_{\rm F}}\sqrt{2\frac{q}{\hbar}B\ell}-\mu_{_{\rm B}}B)}\Big]
\end{eqnarray}
With $\lj N\rj =\beta^{-1}\frac{\p\ln Z}{\p\mu}=-\beta^{-1} \frac{\p\Gamma}{\p\mu} $, we have
\begin{eqnarray}
\lj N\rj_\xi&=&\Delta_{_{\rm L}}\Big[\frac{1}{1+e^{\beta(\mu_{_{\rm B}}B-\mu)}}
+2\sum_{\ell=1}^\infty\frac{1}{1+e^{\beta(\hbar v_{_{\rm F}}\sqrt{2\frac{q}{\hbar}B\ell}+\mu_{_{\rm B}}B-\mu)}}
+2\sum_{\ell=1}^{L_1}\frac{1}{1+e^{\beta(-\hbar v_{_{\rm F}}\sqrt{2\frac{q}{\hbar}B\ell}+\mu_{_{\rm B}}B-\mu)}}\Big]
\end{eqnarray}
Similarly
\begin{eqnarray}
\lj N\rj_\eta&=&\Delta_{_{\rm L}}\Big[\frac{1}{1+e^{\beta(-\mu_{_{\rm B}}B-\mu)}}
+2\sum_{\ell=1}^\infty\frac{1}{1+e^{\beta(\hbar v_{_{\rm F}}\sqrt{2\frac{q}{\hbar}B\ell}-\mu_{_{\rm B}}B-\mu)}}
+2\sum_{\ell=1}^{L_2}\frac{1}{1+e^{\beta(-\hbar v_{_{\rm F}}\sqrt{2\frac{q}{\hbar}B\ell}-\mu_{_{\rm B}}B-\mu)}}\Big]
\end{eqnarray}
At zero temperature, $\mu=0$, and we have
\beq
\lj N\rj=\lj N\rj_\xi+\lj N\rj_\eta
=\Delta_{_{\rm L}}[1+2(L_1+L_2)]
\eeq
and the manetization
\beq
M=\Delta_{_{\rm L}}\mu_{_{\rm B}}(2L_1-2L_2-1)
\eeq
Since $L_1=L_2, L_2+1$, we have $M=\pm \Delta_{_{\rm L}}\mu_{_{\rm B}}$.\\
\section{Perpendicular magnetic and electric fields} 
\subsection{Constant fields: $|\bE|<<\Fermiv |\bB|$}
In this case, we can use the 2+1 Lorentz invariance and the results of previous section. 
Suppose system $\Sigma\yp$ is moving at velocity $\bv$ relative to system $\Sigma$. For $\bv=(v,0)$,
the SO(1,2) transformation is
\beq
A\yp_\mu=\Lambda_\mu\,^\nu A_\nu
\eeq
\beq
\Lambda_\mu\,^\nu=
\left(\begin{array}{ccc}
        \gamma & -\beta\gamma & 0 \\
        -\beta\gamma & \gamma & 0 \\
        0&0&1
      \end{array}\right)
\eeq
where $\beta=\frac{v}{v_{_{\rm F}}}, \gamma=\frac{1}{\sqrt{1-\beta^2}}$
.
\beq
F_{\mu\nu}=\left(\begin{array}{ccc}
        0 & E_x/v_{_{\rm F}} & E_y/v_{_{\rm F}} \\
       -E_x/v_{_{\rm F}} & 0 & -B_z \\
        -E_y/v_{_{\rm F}}&B_z&0
      \end{array}\right)
\eeq
The electromagnetic fields are related by
\beq
\bE\yp_{\parallel}=\bE_{\parallel},\hskip 1in 
\bE\yp_{\perp}=\gamma(\bE+\bv\times \bB)_\perp
\eeq
\beq
\bB\yp_{\parallel}=\bB_{\parallel},\hskip 1in 
\bB\yp_{\perp}=\gamma(\bB-\frac{\bv}{v_{_{\rm F}}^2}\times \bE)_\perp
\eeq
Suppose we have in $\Sigma: \bE=0,\bB=(0,0,B)$, then in $\Sigma\yp$.
\beq
\bE\yp_{\parallel}=0,\hskip 1in 
\bE\yp_{\perp}=\gamma\bv\times \bB
\eeq
\beq
\bB\yp_{\parallel}=\bB_{\parallel},\hskip 1in 
\bB\yp_{\perp}=\gamma\bB_\perp
\eeq
We want $\bE\yp=(0,E\yp_y,0), \bB\yp=(0,0,B\yp)$. Hence we can choose $\bv=(v,0)$. In this case.
\beq
E\yp_y=-\gamma vB\hskip 1in
B\yp_z=\gamma B
\eeq
So we have
\beq
v=-\frac{E\yp_y}{B\yp_z}
\eeq
So in $\Sigma$
\beq
j^{0\prime}=\gamma j^0, j^{x\prime}=-\gamma\beta j^0=\gamma\frac{E\yp_y}{v_{_{\rm F}}B\yp_z}j^0
\eeq
Note that 
\beq
j^\mu=q(v_{_{\rm F}}\rho,\bj)
\eeq
The Hall conductance is
\beq
\sigma\yp_{xy}=\frac{\gamma}{v_{_{\rm F}}B\yp_z}j^0=\frac{q^2}{h}(2L+1)
\eeq
if $-\Bohrmu B$ is filled or
\beq
\sigma\yp_{xy}=\frac{\gamma}{v_{_{\rm F}}B\yp_z}j^0=\frac{q^2}{h}2L
\eeq
if $-\Bohrmu B$ is not filled,
where $L=L_1+L_2$. The physics near $\bKm$ makes equal contribution, hence
\beq
\sigma\yp_{xy}=\frac{\gamma}{v_{_{\rm F}}B\yp_z}j^0
=\left\{\begin{array}{cl}
    \frac{4q^2}{h}(L+\frac{1}{2}), & -\Bohrmu B \text {  is  filled}\\
    \frac{4q^2}{h}L, & -\Bohrmu B \text{  is not filled}
       \end{array}\right.
\eeq
The above SO(1,2) transformation breaks down when $v\geq v_{_{\rm F}}$, which means when
$|E\yp_y|\geq v_{_{\rm F}} |B\yp_z|$. The magnetic moment current
\beq
j^\mu_s=\mu_{_{\rm B}}(v_{_{\rm F}}(\rho_\xi-\rho_\eta),\bj_\xi-\bj_\eta)
\eeq
we then have
\beq
j^{x\prime}_s=(-\gamma\beta)v_{_{\rm F}}M=\mu_{_{\rm B}}\frac{qE\yp_y}{h}
\eeq
So far the sequence of Hall conductance $\sigma_{xy}=\frac{4q^2}{h}(L+\frac{1}{2})$ has been observed, we here predict the existence of the sequence $\sigma_{xy}=\frac{4q^2}{h}L$  and the recently observed $L=0,4$ are just part of this sequence\cite{Jiang:2007}. According to above analysis, the filling of Zeeman levels makes the difference.
\subsection{General orthogonal constant $\bB$ and $\bE$}
When the condition $|\bE|<<\Fermiv |\bB|$ does not hold, we need to solve the Dirac equation directly. 
Let $\bA=(-yB,0), \varphi=yE$. The second order Dirac equation reads
\beq
\Big[-D_iD_i+\frac{q}{\hbar}B\sigma^z+\frac{iqE}{\hbar\Fermiv}\sigma^y
-\frac{(qEy-K+\Bohrmu B)^2}{\hbar^2\Fermiv^2}\Big]\xi=0
\eeq
Denoting $\vep=\frac{qE}{\hbar\Fermiv},\kappa= \frac{K-\Bohrmu B}{\hbar\Fermiv},$
\beq
\Big[-D_iD_i+2\pi\Delta_{_{\rm L}}\sigma^z+i\vep\sigma^y
-(\vep y-\kappa)^2\Big]\xi=0
\eeq
We can diagonalize the part $2\pi\Delta_{_{\rm L}}\sigma^z+i\vep\sigma^y$ by Jordan decomposition (a similarity transformation). 
\beq
2\pi\Delta_{_{\rm L}}\sigma^z+i\vep\sigma^y=SJS^{-1}
\eeq
where
\beq
S=\left(
      \begin{array}{cc}
        \frac{1}{\vep}(-2\pi\Delta_{_{\rm L}}+\sqrt{4\pi^2\Delta_{_{\rm L}}^2-\vep^2}) & \frac{1}{\vep}(-2\pi\Delta_{_{\rm L}}-\sqrt{4\pi^2\Delta_{_{\rm L}}^2-\vep^2}) \\
       1 & 1 \\
      \end{array}
    \right)
\eeq
\beq
J=\left(
      \begin{array}{cc}
        -\sqrt{4\pi^2\Delta_{_{\rm L}}^2-\vep^2} & 0 \\
       0 & \sqrt{4\pi^2\Delta_{_{\rm L}}^2-\vep^2} \\
      \end{array}
    \right):=\left(
      \begin{array}{cc}
        J_1 & 0 \\
       0 & J_2 \\
      \end{array}
    \right)
\eeq
Let 
\beq
\xi=S\Psi
\eeq
Then
\beq
\Big[-D_iD_i+J
-(\vep y-\kappa)^2\Big]\Psi=0
\eeq
\beq
\Big[-\p_y^2-\p_x^2-\frac{2iqB}{\hbar}y\p_x+\frac{q^2B^2}{\hbar^2}y^2+J
-(\vep y-\kappa)^2\Big]\Psi=0
\eeq
\beq
\Big[-\p_y^2-\p_x^2-4\pi\Delta_{_{\rm L}}yi\p_x+4\pi^2\Delta_{_{\rm L}}^2y^2+J
-(\vep y-\kappa)^2\Big]\Psi=0
\eeq
Let
\beq
\Psi=e^{ip_xx/\hbar}f(y)
\eeq
Denoting $k_x=p_x/\hbar$.
\beq
\Big[-\p_y^2+k_x^2+4\pi\Delta_{_{\rm L}}yk_x+4\pi^2\Delta_{_{\rm L}}^2y^2+J
-(\vep y-\kappa)^2\Big]f=0
\eeq
Let
\beq
\alpha=4\pi^2\Delta_{_{\rm L}}^2-\vep^2\hskip 1in 
b=4\pi\Delta_{_{\rm L}}k_x+2\vep \kappa \hskip 1in 
c=k_x^2-\kappa^2+J_1
\eeq
\beq
(-\p_y^2+\alpha y^2+by+c)f_1=0
\eeq
Let 
\beq
u=y+\frac{b}{2\alpha}
\eeq
then
\beq
(-\p_u^2+\alpha u^2+c-\frac{b^2}{4\alpha})f_1=0
\eeq
When $\alpha>0$, this is a harmonic problem while for $\alpha<0$, $J_1$ imaginary. Similar issue exists in \cite{Canuto:1969}, thereof $\beta$ can be larger than 1, hence $\gamma$ is imaginary. The critical value of $B$ is $B_c=E/\Fermiv$. For $B< B_c$, the solution should be parabolic cylinder functions, as in the case of \cite{Su:1984}.\\
\indent For positive $\alpha$, the problem is harmonic oscillator with $m=\hbar^2/2, \omega=2\sqrt{\alpha}/\hbar$. Hence we have
\beq
b^2-4\alpha c=4\alpha\hbar\omega(n+1/2)
\eeq
We have
\begin{eqnarray}
\kappa
&=&\frac{-  \vep  k_x \pm \sqrt{\alpha J_{1,2}+\alpha^{3/2}  +2 \alpha^{3/2} n }}{2\pi \Delta_{_{\rm L}}}
\end{eqnarray}
For $f_2, J_2=\sqrt{\alpha}$, we have
\beq
\kappa=\frac{-  \vep  k_x \pm \sqrt{2 \alpha^{3/2} (n+1) }}{2\pi \Delta_{_{\rm L}}}
\eeq
for $f_1, J_1=-\sqrt{\alpha}$, we have
\beq
\kappa=\frac{-  \vep  k_x \pm \sqrt{2 \alpha^{3/2}n }}{2\pi \Delta_{_{\rm L}}}
\eeq
Now the energy levels depend on $k_x$ and Landau degeneracy is removed partially. 
\beq
E_\pm(k_x,n)=\Bohrmu B+\hbar\Fermiv\frac{-  \vep  k_x \pm \sqrt{2 \alpha^{3/2} n }}{2\pi \Delta_{_{\rm L}}}
\eeq
\subsection{External monochromatic electromagnetic wave}
Consider an external vector potential
\beq
A^\mu=A^\mu_0e^{i(\bk\cdot\br-\omega t)}
\eeq
which represents a monochromatic plane wave and satisfies Lorentz condition 
 \beq
 \p_\mu A^\mu=0
 \eeq
 . $k=\frac{\omega}{c}$. Notice that the plane wave is of speed $c$, not $\Fermiv$. 
  Hence 
\beq
\hat{\bk}\times \bE_0=c\bB_0
\eeq
then $E_0>> \Fermiv B_0$. 
$k^\mu=(\frac{\omega}{\Fermiv},\bk), \bk=|\bk|\hat{\bk}, x^\mu=(\Fermiv t, \bx)$. Note $k^\mu k_\mu=\frac{\omega^2}{\Fermiv^2}-|\bk|^2\not=0$. This is a major difference between graphene system and real Dirac sysytem in applied electromagnetic wave.
We ignore the Zeeman term at first.
\beq
i\hbar \rD \xi=0
\eeq
 As in \cite{LL:1982}, denoting $\phi=\bk\cdot\br-\omega t$ and $A_\mu=A_\mu(\phi)$. Hence
 \beq
\p_\mu A^\mu=k^\mu A'_\mu=0
\eeq 
Or
\beq
k\cdot A:=k_\mu A^\mu=0
\eeq
The filed tensor
\beq
F_{\mu\nu}=k_\mu A'_\nu-k_\nu A'_\mu
\eeq
The second-order equation is
\beq
(D^\mu D_\mu+\frac{1}{2}\gamma^\mu\gamma^\nu i\frac{q}{\hbar}F_{\mu\nu})\xi=0
\eeq
Using
\beq
D_\mu D^\mu\xi=(\p^2+2i\frac{q}{\hbar}A^\mu\p_\mu -\frac{q^2}{\hbar^2}A^2)\xi
\eeq
and
\beq
[\gamma^\mu,\gamma^\nu]k_\mu A'_\nu=2(\gamma\cdot k)(\gamma\cdot A')
\eeq
we have
\beq
\Big[\p^2+2i\frac{q}{\hbar}A^\mu\p_\mu -\frac{q^2}{\hbar^2}A^2+i\frac{q}{\hbar}(\gamma\cdot k)(\gamma\cdot A')\Big]\xi=0
\eeq
We seek a solution of this equation, a la the original Volkow ansatz \cite{Volkow:1935}\cite{LL:1982}
\beq
\xi=e^{-ip\cdot x}F(\phi)
\eeq
then ($A'=iA$)
\beq
k^2F''-p^2F-2ik\cdot p F'+\frac{2q}{\hbar}A\cdot p F-\frac{q^2}{\hbar^2}A^2F-\frac{q}{\hbar}(\gamma\cdot k)(\gamma\cdot A)F=0 \label{Feq}
\eeq
We can impose (since we can always add to $p$ a multiple of $k$ to meet this condition yet the the functional form of $\xi$ remains.)
$
p^2=0
$
or some other condition
to simplify the equation (\ref{Feq}), but it will always be a second-order equation. The reason is that here $k^\mu k_\mu\not=0$, as distinguishes graphene system from the conventional relativistic electrons. \\
\section{Summary}
\indent To summarize, we discussed the Hamiltonian and energy levels of graphene in general constant external electric and magnetic fields. The systems is not SO(1,2) Lorentz invariant when Zeeman energy is taken into account. But when the magnetic field is strong enough, SO(1,2) Lorentz invariance is well preseved. Employing the symmetry, we predicted a sequence $\sigma_{xy}=\frac{4e^2}{h}L$ and explain the recently observed Hall conductance
$\sigma_{xy}=\frac{4e^2}{h}L, L=0,4$, which is an indication that the Zeeman levels are not filled at zero temperature. 
The second-order Dirac equation is derived when the applied field is a monochromatic electromagnetic wave and the difference between graphene sysytem and standard relativistic electrons in this case is revealed.

\begin{acknowledgments}
 M. M. was partially supported by NSF Grant No. DMR-0804805.
\end{acknowledgments}

\end{document}